\documentclass[
	amsmath,
	amssymb,
	a4paper,
	pre,		
	reprint,	
	fleqn,
	showpacs,
	floatfix,
	superscriptaddress,
        twocolumn
]{revtex4-1}

\usepackage[utf8]{inputenc}
\usepackage[T1]{fontenc}
\usepackage{microtype}

\usepackage[lining]{libertine}
\usepackage[libertine,slantedGreek,bigdelims,vvarbb]{newtxmath}
\useosf 
\renewcommand{\Delta}{\upDelta}

\usepackage{tabularx}
\usepackage{booktabs}
\usepackage{multirow}

\usepackage{bm}
\usepackage{amssymb}
\usepackage{braket}

\usepackage{kbordermatrix}
	\renewcommand{\kbldelim}{(}
	\renewcommand{\kbrdelim}{)}
\usepackage{graphicx}
\usepackage{dcolumn}
\usepackage[caption=false]{subfig}
\usepackage{xcolor}

\usepackage[version=3]{mhchem}
\usepackage{tikz}
\usetikzlibrary{matrix,positioning,decorations.pathreplacing}

\usepackage{hyperref}

\newcommand{\myTitle}{Thermodynamic Efficiency in Dissipative Chemistry}

\newcommand\unilu{\affiliation{Complex Systems and Statistical Mechanics, Physics and Materials Science Research Unit, University of Luxembourg, L-1511 Luxembourg, G.D.~Luxembourg}}
\newcommand\ias{\affiliation{\emph{Present Address: } The Simons Center for Systems Biology, School of Natural Sciences, Institute for Advanced Study, Princeton, 08540 New Jersey, U.S.A.}}
\newcommand{\emanuele}{Emanuele Penocchio}
\newcommand{\riccardo}{Riccardo Rao}
\newcommand{\massimiliano}{Massimiliano Esposito*}

\newcommand{\myFunding}{
	This work was funded by the \emph{Luxembourg National Research Fund} (AFR PhD Grant 2014-2, No.~9114110) and the \emph{European Research Council} project NanoThermo (ERC-2015-CoG Agreement No.~681456). 
}

\newcommand{\der}[2]{\frac{\mathrm{d}{#1}}{\mathrm{d}{#2}}}

\newcommand{\dt}{\mathrm{d}_{t}}

\newcommand{\at}[2]{\left.{#1}\right|_{#2}}

\newcommand{\transpose}{^{\mathsf{T}}}

\newcommand{\ep}{\Sigma}
\newcommand{\epr}{\dot{\ep}}

\DeclareMathOperator{\diag}{diag}

\definecolor{webgreen}{rgb}{0,.5,0}
\definecolor{webbrown}{rgb}{.6,0,0}
\definecolor{grigio}{rgb}{.85,.85,.85} 
\definecolor{RoyalBlue}{rgb}{0.0, 0.14, 0.4}
\definecolor{skyblue3}{rgb}{0.13,0.29,0.53}
\definecolor{aluminium5}{rgb}{0.33,0.34,0.32}

\newcommand{\greyt}[1]{\textcolor{aluminium5}{#1}}

\hypersetup{%
    colorlinks=true, linktocpage=true, pdfstartpage=1, pdfstartview=FitV,%
    breaklinks=true, pdfpagemode=UseNone, pageanchor=true, pdfpagemode=UseOutlines,%
    plainpages=false, bookmarksnumbered, bookmarksopen=true, bookmarksopenlevel=1,%
    hypertexnames=true, pdfhighlight=/O,
    urlcolor=webbrown, linkcolor=RoyalBlue, citecolor=webgreen, 
    pdftitle={\myTitle},%
    pdfauthor={\emanuele},%
    pdfsubject={},%
    pdfkeywords={},%
    pdfcreator={pdfLaTeX},%
    pdfproducer={LaTeX REVTeX}%
}

\renewcommand{\thesection}{\Roman{section}}

\renewcommand{\L}{\mathrm{L}}

\newcommand{\F}{\mathrm{F}}
\newcommand{\W}{\mathrm{W}}

\newcommand{\M}{\mathrm{M}}
\newcommand{\Ms}{\mathrm{M}^{\ast}}
\newcommand{\A}{\mathrm{A}_{2}}
\newcommand{\As}{\mathrm{A}_{2}^{\ast}}

\newcommand{\cF}{[\F]}
\newcommand{\cW}{[\W]}

\newcommand{\cM}{[\M]}
\newcommand{\cMs}{[\Ms]}
\newcommand{\cA}{[\A]}
\newcommand{\cAs}{[\As]}

\newcommand{\matS}{\mathbb{S}}
\newcommand{\matSX}{\mathbb{S}^{\mathrm{X}}}
\newcommand{\matSY}{\mathbb{S}^{\mathrm{Y}}}

\begin{document}

\title{\myTitle}

\author{\emanuele}
\unilu
\author{\riccardo}
\unilu
\ias
\author{\massimiliano}
\unilu


\begin{abstract}
        Chemical processes in closed systems are poorly controllable since they always relax to equilibrium. 
        Living systems avoid this fate and give rise to a much richer diversity of phenomena by operating under nonequilibrium conditions~\cite{zwa15,grz16,hess17}. 
        Recent experiments in dissipative self-assembly also demonstrated that by opening reaction vessels and steering certain concentrations, an ocean of opportunities for artificial synthesis and energy storage emerges~\cite{otto15,ros17,rag18}.
        To navigate it, thermodynamic notions of energy, work and dissipation must be established for these open chemical systems. 
        Here, we do so by building upon recent theoretical advances in nonequilibrium statistical physics~\cite{seifert12,ciliberto17}.
        As a central outcome, we show how to quantify the efficiency of such chemical operations  
        and lay the foundation for performance analysis of any dissipative chemical process.
\end{abstract}

\maketitle

Traditional chemical thermodynamics deals with closed systems, which always evolve towards equilibrium.
At equilibrium, all reaction currents --- defined as the forward reaction fluxes minus the backwards ($J_\rho = J_{+\rho} - J_{-\rho}$, where $\rho$ labels the reactions) --- eventually vanish.
The first thermodynamic description of nonequilibrium chemical processes was achieved by the Brussels school founded by de~Donder and perpetuated by Prigogine~\cite{prigogine54,prigogine67}, but they focused on few reactions close to equilibrium in the so-called linear regime.
However, processes such as dissipative self-assembly are open chemical reaction networks (CRN) involving many reactions operating far away from equilibrium.
The openness arises from the presence of one or more \emph{chemostats}, i.e. particle reservoirs coupled with the system which externally control the concentrations of some species --- just like thermostats control temperatures --- and allow for matter exchanges.
Open CRN can then be thought of as thermodynamic machines powered by chemostats.
Two operating regimes may be distinguished, reminiscent of stroke and steady-state engines.
In the first, work is used to induce a time-dependent change in the species abundances that could never be reached at equilibrium.
An example could be the accumulation of a large amount of molecules with a high free energy content as in dissipative self-assembly, or the depletion of some undesired species as in metabolite repair~\cite{linster13}.
In the second, work is used to maintain the system in a nonequilibrium stationary state which continuously transduces an input work into useful output work.
Currently no framework exists to asses how efficient and powerful such chemical engines can be.
We provide one grounded in the recently established nonequilibrium thermodynamics of CRN~\cite{rao16:crnThermo,falasco18}, which was born from the combination of state-of-the-art statistical mechanics~\cite{jarzynski11,seifert12,qian12a,qian12b,vandenbroeck15} and mathematical CRN theory~\cite{horn72,feinberg72}.
Rigorous concepts of free energy, chemical work and dissipation valid far from equilibrium reveal crucial.  
They provide the basis for thermodynamically meaningful definitions of efficiencies and optimal performance in the different operating regimes. 
In the following, energy storage (ES) and dissipative synthesis (DS) will be analyzed as models epitomizing the first and the second operating regime, respectively, but our findings apply to any dissipative chemical process.  
\bigskip


\begin{figure}[b!]
    \includegraphics[width=.4\textwidth]{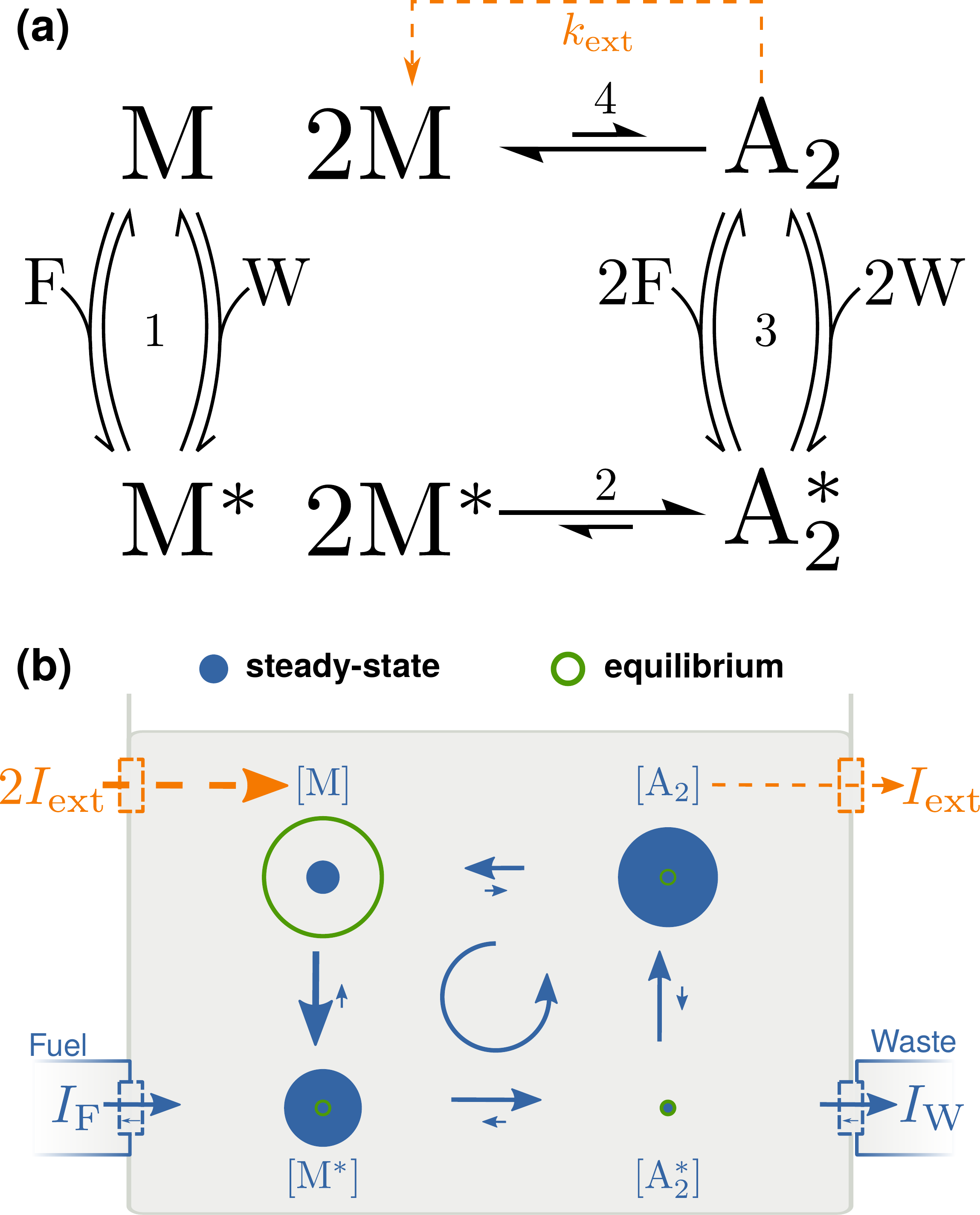}
    \centering
    \caption{\textbf{| Model for energy storage and dissipative synthesis.}
                Without (resp. with) the orange dashed transition, the chemical reaction network models energy storage (resp. dissipative synthesis). 
                The high energy species $\A$ is at low concentration at equilibrium.
                Powering the system by chemostatting fuel ($\F$) and waste ($\W$) species boosts the formation of $\A$ out of the monomer $\M$ via the activated species $\Ms$ and $\As$.
                \textbf{a,} The chemical reaction network (forward fluxes are defined counter-clockwise).
		\textbf{b,} Sketch of concentrations distributions (proportional to radii).
}
\label{FIG_scheme}
\end{figure}

\begin{figure}[b]
	\includegraphics[width=.45\textwidth]{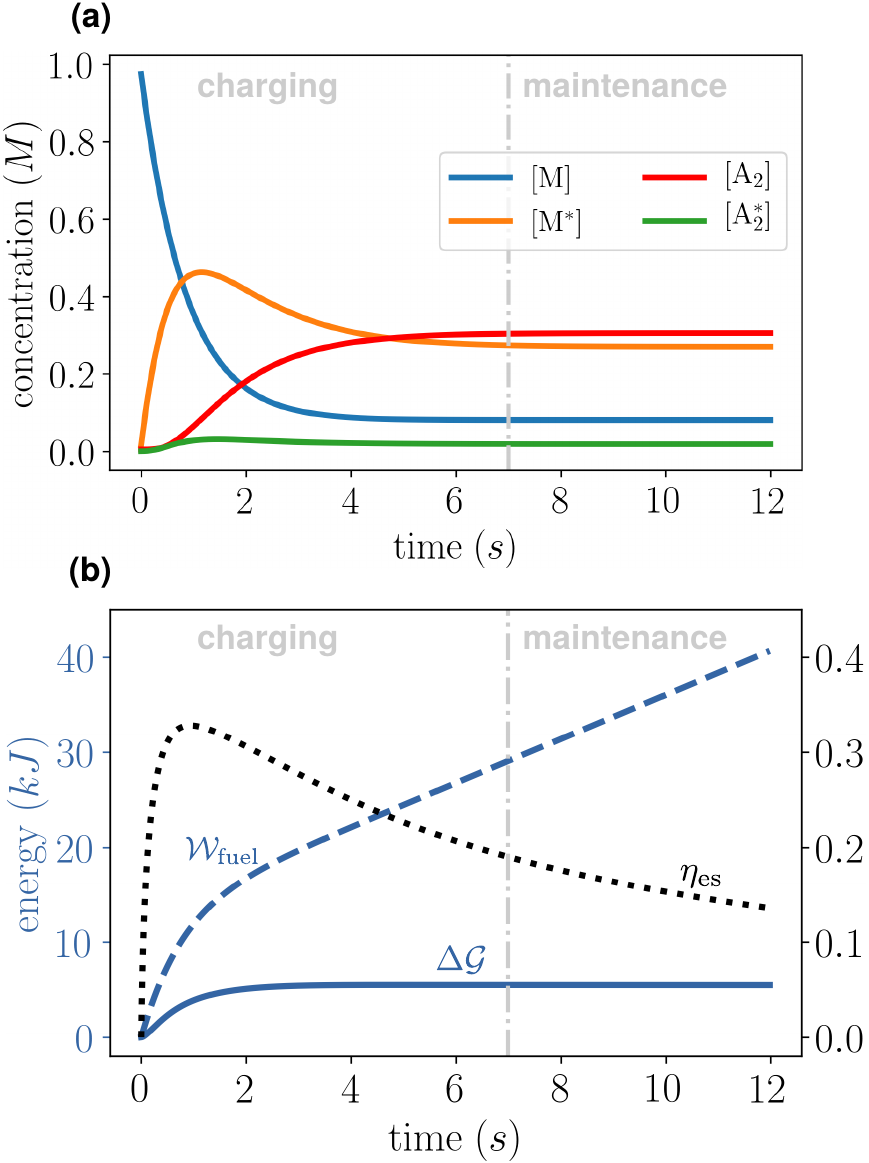}
	\centering
	\caption{$\bf{|}$ \textbf{Dynamics of energy storage}.
        The system is initially prepared at thermodynamic equilibrium where $\cM_{\mathrm{eq}} \gg \cA_{\mathrm{eq}}$.
        At time $t=0$, the chemical potential difference between fuel and waste is turned on at $\mathcal{F}_{\mathrm{fuel}} = 7.5 \cdot RT$ and drives the system away from equilibrium.
        After a transient (charging phase), the system eventually settles into a nonequilibrium steady state (maintenance phase). 
        \textbf{a,} Species abundances. \textbf{b,} Energy stored, work and efficiency (right axis, adimensional units). 
        Kinetic constants ($\{k_{\pm\rho}\}$) and chemical potentials used for simulations are given in Supplementary Information.
}
\label{FIG_ES}
\end{figure}

\begin{figure}[t]
	\centering
	\includegraphics[width=.4\textwidth]{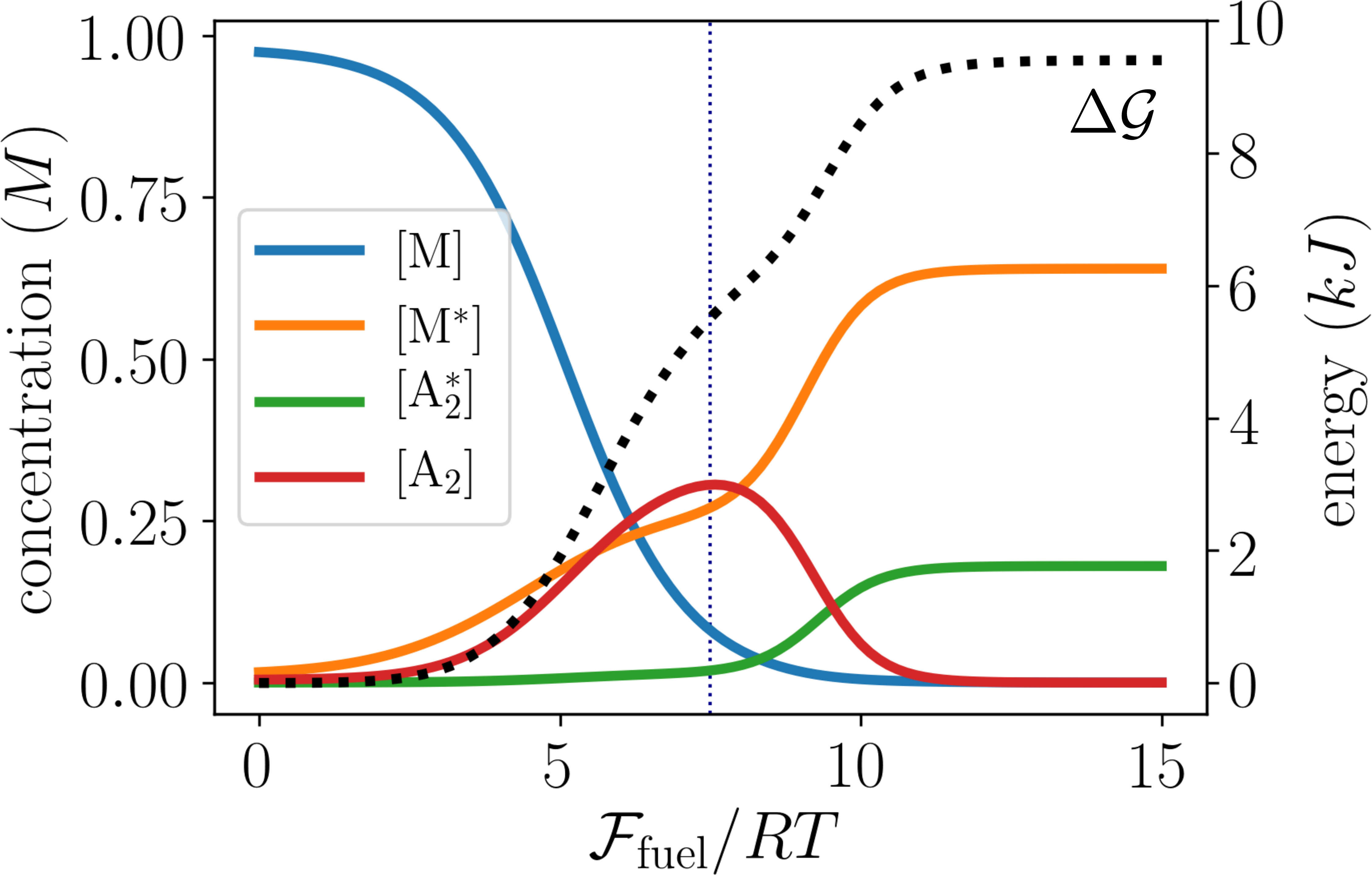}
	\caption{$\bf{|}$ \textbf{Maintenance phase of energy storage}.
            Stationary concentrations and free energy difference from equilibrium in the maintenance phase of energy storage as a function of the chemical potential difference between fuel and waste.
	    The vertical dotted line denotes the value $\mathcal{F}_\mathrm{fuel} = 7.5 \cdot RT$ used to study the charging phase in Fig.~\ref{FIG_ES}.
        }
	\label{fig:harvesting}
\end{figure}

In \emph{energy storage}, an open CRN initially at equilibrium with high concentrations of low-energy molecules and low concentrations of high-energy ones, is brought out of equilibrium with the aim to increase the concentrations of the high-energy species.
This process is reminiscent of charging a capacitor via the coupling to a voltage generator.
In the context of supramolecular chemistry, the concept of ES was proposed by Ragazzon and Prins~\cite{rag18}. 
An insightful model capturing its main features is described in Fig.~\ref{FIG_scheme}. 
Its thermodynamic analysis, detailed in the Supplementary Information, will now be outlined.  
The accumulation of the high energy species $\A$ is enabled when chemostats set a positive chemical potential difference of fuel and waste, i.e. $\mathcal{F}_{\mathrm{fuel}} = \mu_\F - \mu_\W >0$.
This implies the injection of $\F$ molecules at a rate $I_{\F}$ and the extraction of $\W$ at rate $I_{\W}$.
The resulting power (i.e., work per unit of time) performed on the system by the fueling mechanism is $\dot{\mathcal{W}}_{\mathrm{fuel}} = I_\F \, \mathcal{F}_{\mathrm{fuel}}$~\cite{rao16:crnThermo,falasco18}.
The proper way to quantify the energy content of an open CRN is via its nonequilibrium free energy $\mathcal{G}$.
During the charging process, only part of the work, namely $\Delta \mathcal{G}$, is dedicated to accumulate the high energy species $\A$~\cite{rag18} and is stored as free energy in the system.
The remaining fraction, namely $T \Sigma$, is dissipated according to the second law of thermodynamics
\begin{align}
	\mathcal{W}_{\mathrm{fuel}} = \Delta \mathcal{G} + T \Sigma\, ,
	\label{es_work}
\end{align}
where $T$ is temperature and $\Sigma \geq 0$ the entropy production which only vanishes at equilibrium. 
The thermodynamic efficiency of the ES process is thus the ratio 
\begin{align}
	\eta_{\mathrm{es}} = \frac{\Delta\mathcal{G}}{\mathcal{W}_{\mathrm{fuel}}} = 1 - \frac{T\Sigma}{\mathcal{W}_{\mathrm{fuel}}} \, .
	\label{es_eta}
\end{align}
Eq.~\ref{es_work} has been used to derive the second inequality.
We emphasize that each of these contributions has an explicit expression in terms of concentrations and rate constants, see Supplementary Information.
For instance, the energy stored at any time with respect to equilibrium is given by the expression
\begin{align}
    \Delta \mathcal{G} = RT \sum_{\substack{X=\M,\\\Ms,\As,\A}} \left[ [X] \ln \frac{[X]}{[X]_\mathrm{eq}} - [X] + [X]_\mathrm{eq} \right] \geq 0\, ,
	\label{es_gibbs}
\end{align}
which can be recognized as the information measure called relative entropy \cite{CoverThomas06}.
Since $\Delta \mathcal{G}$ has to be positive in ES, the second law implies that work has to be positive (done on the system).
It also ensures that $\eta_{\mathrm{es}}$ is bounded between zero and one.

We simulated an ES process and plotted the dynamics of concentrations as well as efficiency and its contributions in Fig.~\ref{FIG_ES}.
The process can be divided into a \emph{charging} and a \emph{maintenance} phase.
During the former, the system energy grows ($\dt \mathcal{G} > 0$) in a way which correlates with the accumulation of the high energy species $\A$.
\begin{figure*}[t]
	\includegraphics[width=.95\textwidth]{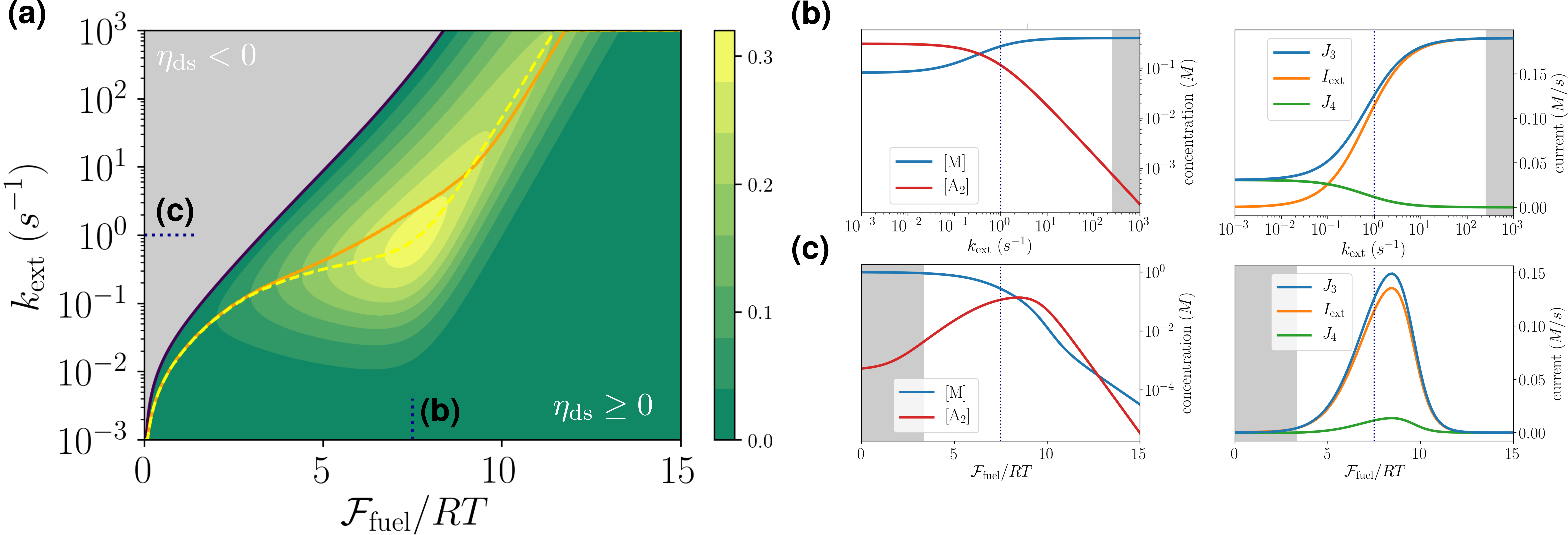}
	\centering
	\caption{
                \textbf{| Performance of dissipative synthesis.}
                \textbf{a,} Efficiency ($\eta_\mathrm{ds}$) of dissipative synthesis as function of $\mathcal{F}_\mathrm{fuel}$ and $k_\mathrm{ext}$.
                Regions of operating regimes that do not perform dissipative synthesis are colored in gray.
                The yellow dashed (resp. orange solid) line denotes the maximum of $\eta_\mathrm{ds}$ (resp. $-\dot{W}_\mathrm{ext}$) versus $k_\mathrm{ext}$, while the black solid line corresponds to $\eta_\mathrm{ds}=0$.
                \textbf{b,} (resp. \textbf{c,}) Currents and concentrations as a function of $k_\mathrm{ext}$ (resp. $\mathcal{F}_\mathrm{fuel}$) for $\mathcal{F}_\mathrm{fuel}=7.5 \cdot RT$ (resp. $k_\mathrm{ext}=1 \, s^{-1}$) denoted by blue dotted ticks in plot \textbf{a}.
                Kinetic constants and standard chemical potentials are the same as for ES analysis (see Supplementary Information).
                Note that $J_3 = I_\mathrm{ext} + J_\mathrm{4}$ always holds, where $J_3 = J_{3\mathrm{F}} + J_{3\mathrm{W}}$ is the net current flowing from $\As$ to $\A$.
	}
	\label{FIG_DS}
\end{figure*}
The process can be quite efficient as a significant portion of the work is converted into free energy. 
However, in the maintenance phase, the system reaches a nonequilibrium steady state. 
The efficiency drops towards zero (proportional to the inverse time) as the entire work is being spent to preserve the energy previously accumulated ($\dt \mathcal{G} \simeq 0$).
Figure~\ref{fig:harvesting} focuses on the maintenance phase for different values of $\mathcal{F}_{\mathrm{fuel}}$.
It shows that by driving the system away from equilibrium, one can reach species abundances which are very different with respect to the equilibrium ones.
It also shows that the accumulation of free energy does not necessarily coincide with an increase in concentration of the most energetic species $\A$.
Indeed, while at low values of $\mathcal{F}_{\mathrm{fuel}}$ the accumulation of $\mathcal{G}$ correlates with $\cA$, beyond a critical point, $\A$ starts to be depleted while energy continues getting stored by further moving away the concentration distribution from equilibrium.

As we have seen, the crucial part of energy storage is the charging phase, as the maintenance phase is purely dissipative and burns chemical work without any energy gain.  
In order to make use of the energy accumulated during the charging phase, a mechanism extracting the energetic species from the system must be introduced.     
This complementary but distinct working regime of an open CRN will now be considered. 
\bigskip


In \emph{dissipative synthesis}, an energetic species which accumulates thanks to a fueling process is continuously extracted from a system in a nonequilibrium steady state. 
One may consider for instance processes where the product either evaporates, precipitates or undergoes other fast transformations while being rapidly replaced by reactants.
By building upon the above ES scheme, a simple way to model DS is to add an ideal extraction/injection mechanism to the CRN (orange dashed arrows in Fig.~\ref{FIG_scheme}).
This mechanism removes the assembled molecule $\A$ and renews two $\M$ molecules at a rate $I_{\mathrm{ext}} = k_{\mathrm{ext}} \cA$.
In doing so, we model the synthesis of molecules that are strongly unfavored at equilibrium, a strategy used by Nature~\cite{des97,how01,hess17} and which may be within reach of supramolecular chemists~\cite{boe10,boe15,sor17}.

From the thermodynamic standpoint detailed in the Supplementary Information, the input power spent by the fueling mechanism, $\dot{\mathcal{W}}_{\mathrm{fuel}} = I_{\F} \, \mathcal{F}_\mathrm{fuel} = I_{\F} \, (\mu_{\F} - \mu_{\W})$, is now not just dissipated as $T\dot{\Sigma}$, but part of it is used to sustain the production of $\A$:
\begin{align}
	\dot{\mathcal{W}}_{\mathrm{fuel}} = - \dot{\mathcal{W}}_{\mathrm{ext}} + T\dot{\Sigma} \, .
	\label{ds_power}
\end{align}
The output power released by the extraction mechanism, $\dot{\mathcal{W}}_{\mathrm{ext}} = I_{\mathrm{ext}} (2 \mu_\mathrm{M} - \mu_{\mathrm{A_2}})$, is negative when DS occurs.
In this context the thermodynamic efficiency is thus given by 
\begin{align}
	\eta_{\mathrm{ds}} = - \frac{\dot{\mathcal{W}}_{\mathrm{ext}}}{\dot{\mathcal{W}}_{\mathrm{fuel}}} = 1 - \frac{T\dot{\Sigma}}{\dot{\mathcal{W}}_{\mathrm{fuel}}} \, ,
	\label{ds_eta}
\end{align}
where Eq.~\ref{ds_power} has been used to derive the second equality.
It is bounded between zero and one when DS occurs.

In Fig.~\ref{FIG_DS}, we simulated DS for various working conditions by varying $k_\mathrm{ext}$ and $\mathcal{F}_\mathrm{fuel}$. 
We start our analysis by considering a given value of $\mathcal{F}_\mathrm{fuel}$. 
As $k_\mathrm{ext}$ is increased, $\eta_{\mathrm{ds}}$ first grows to an optimal value before decreasing towards negative values where the DS regime ends (see Fig.~\ref{FIG_DS}a).
At the same time $I_\mathrm{ext}$ increases until it reaches a plateau. This happens when $k_\mathrm{ext}$ overcomes the ability of the system to sustain high values of $\cA$ (Fig.~\ref{FIG_DS}b).
Eventually the drop in $\cA$ is such that $2 \mu_\mathrm{M} - \mu_{\mathrm{A_2}} > 0$, thus resulting in the loss of the DS regime. 
We now fix $k_\mathrm{ext}$ and increase $\mathcal{F}_\mathrm{fuel}$ (Fig.~\ref{FIG_DS}a and~\ref{FIG_DS}c).
The DS regime starts at a threshold value, when $\cA$ becomes high enough.
After that, both $\cA$ and the efficiency grow to an optimal value before decreasing again.
This time however, the efficiency remains positive as $\cM$ drops together with $\cA$. 
Fig.~\ref{FIG_DS}c shows another important feature. As $\mathcal{F}_\mathrm{fuel}$ is increased, $I_\mathrm{ext}$ first increases too, but eventually reaches a maximum after which it decreases.
This phenomenon is an hallmark of far-from-equilibrium physics which could not happen in a linear regime, namely when $k_\mathrm{ext}$ and $\mathcal{F}_\mathrm{fuel}$ are small.
Remarkably, the global maximum of the efficiency in Fig.~\ref{FIG_DS}a is reached in a region far from equilibrium.
We note that it corresponds to values of $\mathcal{F}_\mathrm{fuel}$ close to the one maximizing $\cA$ in the maintenance phase of ES (see Fig.~\ref{fig:harvesting}) and to values of $k_\mathrm{ext}$ of order one resulting in $I_\mathrm{ext}$ which do not overly deplete $\cA$.
We finally turn to the lines of maximum efficiency and efficiency at maximum power, where the maximization is done with respect to $k_\mathrm{ext}$ at a given $\mathcal{F}_\mathrm{fuel}$ (Fig.~\ref{FIG_DS}a).
Since these two lines typically do not coincide, the study of the tradeoffs is the object of a rich field called finite-time thermodynamics~\cite{benenti17}.
Interestingly, while these two lines cannot coincide in the linear regime (see Supplementary Information), we see that they do intersect far-from-equilibrium, not far from the global maximum of the efficiency.
Our analysis thus allowed us to identify a region of good tradeoff between power and efficiency for the model of DS we introduced.   
In order to emphasize the fact that all the interesting features that we identified in DS occur far-from-equilibrium, we analyze in detail in the Supplementary Information the linear regime of DS.
After identifying the Onsager matrix, we are able to reproduce the entire bottom-left part of Fig.~\ref{FIG_DS}a analytically as well as the behavior of the maximum efficiency and efficiency at maximum power in that region.

Thermodynamics was born from the effort to systematize the performance of steam engines. 
Open chemical reaction networks, which are at the core of recent efforts in artificial synthesis and ubiquitous in living systems, can be seen as chemical engines.
In the spirit of this analogy, in this letter we built a chemical thermodynamic framework which enables us to systematically analyze the performance of two fundamental dissipative chemical processes. 
The first, energy storage, is concerned with the time dependent accumulation of high energy species far from equilibrium and is currently raising significant attention from supramolecular chemists.  
The second, dissipative synthesis, aims at continuously extracting the previously obtained high energy species and provides a simple and insightful instance of energy transduction.       
In doing so, we identified their optimal regimes of operation.
Crucially they lie far-from-equilibrium in regions unreachable using conventional linear regime thermodynamics.   
We emphasize that the methods developed in this paper can in principle be applied to any open chemical reaction network and thus provide the basis for future performance studies and optimal design of dissipative chemistry.   
They are thus destined to play a major role in bioengineering and nanotechnologies.

\begin{acknowledgments}
	\myFunding
\end{acknowledgments}

\begin{widetext}
\appendix

\renewcommand\thefigure{\thesection.\arabic{figure}}    
\renewcommand\thetable{\thesection.\arabic{table}}

\section{Energy Storage}
\label{sec:ES}

\subsection{Dynamics}
\label{sec:ESdynamics}

The evolution in time of the concentrations of the species $\M$, $\Ms$, $\As$, and $\A$ is ruled by the rate equations
\begin{equation}
	\dt
	\underbrace{
		\begin{pmatrix}
			\cM \\
			\cMs \\
			\cAs \\
			\cA
		\end{pmatrix}
	}_{\textstyle [\bm{X}]}
	= 
	\underbrace{
		\begin{pmatrix}
			-1 	& -1	& 0 	& 0  	& 0 	& 2 \\
			1 	& 1		& -2 	& 0  	& 0 	& 0 \\
			0 	& 0		& 1 	& -1  	& -1 	& 0 \\
			0 	& 0		& 0 	& 1  	& 1 	& -1
		\end{pmatrix}
	}_{\textstyle \matSX}
        \cdot
	\underbrace{
		\begin{pmatrix}
			k_{+1\F} \cF \cM - k_{-1\F} \cMs \\
			k_{+1\W} \cW \cM - k_{-1\W} \cMs \\
			k_{+2} \cMs^{2} - k_{-2} \cAs \\
			k_{+3\F} \cAs - k_{-1\F} \cA \cF \\
			k_{+3\W} \cAs - k_{-1\W} \cA \cW \\
			k_{+4} \cA - k_{-2} \cM^{2} \\
		\end{pmatrix}
	}_{\textstyle \bm{J} = \bm{J}_{+} - \bm{J}_{-}}
	\, ,
	\label{eq:ESrateX}
\end{equation}
where $\cF$ and $\cW$ are the concentrations of fuel and waste species.
Since these latter are externally kept constant by the \emph{chemostats}, the balance equations for their concentrations read
\begin{equation}
	\bm{0} = 
	\dt
	\underbrace{
		\begin{pmatrix}
			\cF \\
			\cW 
		\end{pmatrix}
	}_{\textstyle [\bm{Y}]}
	= 
	\underbrace{
		\begin{pmatrix}
			-1 	& 0		& 0 	& 2  	& 0 	& 0 \\
			0 	& -1	& 0 	& 0  	& 2 	& 0
		\end{pmatrix}
	}_{\textstyle \matSY}
        \cdot
	\underbrace{
		\begin{pmatrix}
			k_{+1\F} \cF \cM - k_{-1\F} \cMs \\
			k_{+1\W} \cW \cM - k_{-1\W} \cMs \\
			k_{+2} \cMs^{2} - k_{-2} \cAs \\
			k_{+3\F} \cAs - k_{-1\F} \cA \cF \\
			k_{+3\W} \cAs - k_{-1\W} \cA \cW \\
			k_{+4} \cA - k_{-2} \cM^{2} \\
		\end{pmatrix}
	}_{\textstyle \bm{J} = \bm{J}_{+} - \bm{J}_{-}}
	+
	\underbrace{
		\begin{pmatrix}
			I_{\F} \\
			I_{\W} \\
		\end{pmatrix}
	}_{\textstyle \bm I}
	\, ,
	\label{eq:ESrateY}
\end{equation}
with $I_{\F}$ and $I_{\W}$ denoting the external currents of fuel and waste flowing from the chemostats.
We denote by $X=\M,\Ms,\A,\As$ the internal species, by $Y=\F,\W$ the chemostatted ones, and label by $\rho=1\F,1\W,2,3\F,3\W,4$ all reactions.

\subsection{Thermodynamics}
\label{sec:ESthermo}

We imagine an isothermal, isobaric, and well-stirred ideal dilute solution.
Then, each species is thermodynamically characterized by chemical potentials of the form
\begin{equation}
	\begin{split}
		{\mu}_{X} & = {\mu}^\circ_{X} + RT \ln \frac{[{X}]}{[0]} \, , & 
		{\mu}_{Y} & = {\mu}^\circ_{Y} + RT \ln \frac{[{Y}]}{[0]} \, ,
	\end{split}
	\label{}
\end{equation}
where ${\mu}^\circ_{X}$ and ${\mu}^\circ_{Y}$ are standard-state chemical potentials and $[0]$ is the standard-state concentration.

Dynamics and thermodynamics are related via the hypothesis of local detailed balance, which relates the ratio of rate constants to the differences of standard-state chemical potentials along reactions
\begin{equation}
	RT \ln \frac{k_{+\rho}}{k_{-\rho}} = - {\sum_{X}} \, \mu^\circ_{X} \matS^{X}_{\rho} - {\sum_{Y}} \, \mu^\circ_{Y} \matS^{Y}_{\rho} \, .
	\label{eq:ldb}
\end{equation}
At equilibrium, the thermodynamic forces driving each reaction, also called \emph{affinities}, vanish
\begin{equation}
	A_{\rho}^{\mathrm{eq}} = - {\textstyle\sum_{X}} \mu_{X}^{\mathrm{eq}} \matS^{X}_{\rho} - {\textstyle\sum_{Y}} \mu_{Y}^{\mathrm{eq}} \matS^{Y}_{\rho} = 0 \, ,
	\label{}
\end{equation}
as well as all reaction currents
\begin{equation}
	J_{\rho}^{\mathrm{eq}} = J^{\mathrm{eq}}_{+\rho} - J^{\mathrm{eq}}_{-\rho}  = 0 \, .
	\label{}
\end{equation}

The dissipation of the process is captured by the entropy production (EP) rate, also vanishing at equilibrium
\begin{equation}
	T \epr = RT {\sum_{\rho}} \, J_{\rho} \ln \frac{J_{+\rho}}{J_{-\rho}} \ge 0 \, .
	\label{eq:eprlog}
\end{equation}
Using the rate equations and the local detailed balance, Eq.~\eqref{eq:ldb}, one can rewrite this quantity as
\begin{equation}
	T \epr = - \dt G + \dot{\mathcal{W}}_{\mathrm{chem}} \, ,
	\label{eq:eprCW}
\end{equation}
where
\begin{equation}
	G = {\textstyle \sum_{X}} [X] \left( \mu_{X} - RT \right) + {\textstyle \sum_{Y}} [Y] \left( \mu_{Y} - RT \right)
	\label{}
\end{equation}
is the \emph{Gibbs free energy}, while
\begin{equation}
	\dot{\mathcal{W}}_{\mathrm{chem}} = {\textstyle \sum_{Y}} \mu_{Y} I_{Y} = \mu_{\F} I_{\F} + \mu_{\W} I_{\W}
	\label{eq:ESCW}
\end{equation}
is the \emph{chemical work} per unit time exchanged with the chemostats.

One can also show that if the CRN were closed (fuel and waste not chemostatted) it would relax to equilibrium by minimizing $G$ \cite{rao16:crnThermo}.
Fuel and waste are however chemostatted and we need to identify the conditions for equilibrium in the open CRN.
To do so we preliminary identify the topological properties of the network.

The stoichiometric matrix $\matS \equiv (\matSX,\matSY)^{\mathsf{T}}$ (see  Eqs.~\eqref{eq:ESrateX} and \eqref{eq:ESrateY}) encodes the topological properties of the CRN.
We can access these properties by determining its cokernel, which is spanned by
\begin{align}
	\bm \ell_{\M} & =
	\kbordermatrix{
		& \greyt{\M} & \greyt{\Ms} & \greyt{\As} & \greyt{\A} & \greyt{\F} & \greyt{\W} \\
		& 1 & 1 & 2 & 2 & 0 & 0
	} \, , \label{eq:ESclL} \\
	\bm \ell_{\W} & =
	\kbordermatrix{
		& \greyt{\M} & \greyt{\Ms} & \greyt{\As} & \greyt{\A} & \greyt{\F} & \greyt{\W} \\
		& 0 & 1 & 2 & 0 & 1 & 1
	} \, .  \label{eq:ESclMMbGauge}
\end{align}
The first of these vectors identifies a conserved quantity
\begin{align}
	&L_{\M} = \bm \ell_{\L} \cdot
	\begin{pmatrix}
		[\bm{X}] \\
		[\bm{Y}]
	\end{pmatrix}
	= \cM + \cMs + 2 \cAs + 2 \cA \, , \nonumber \\
        &\dt L_{\M} = 0
	\label{eq:EScl}
\end{align}
which is proved using the rate equations Eq.~\eqref{eq:ESrateX} and Eq.~\eqref{eq:ESrateY}.

The second vector identifies what we call a \emph{broken} conserved quantity
\begin{equation}
	L_{\W} = \bm \ell_{\W} \cdot
	\begin{pmatrix}
		[\bm{X}] \\
		[\bm{Y}]
	\end{pmatrix}
	= \cMs + 2 \cAs + \cF + \cW  \, .
\end{equation}
Using again the rate equations, it can be shown that
\begin{equation}
	\dt L_{\W} := I_{\F} + I_{\W} \, .
	\label{eq:EScqW}
\end{equation}
Namely, $L_{\W}$ changes only due to the exchange of fuel and waste with the chemostats.
If the CRN were closed, $L_{\W}$ would be constant.
Using Eq.~\eqref{eq:EScqW}, we can rewrite the entropy production in Eq.~\eqref{eq:eprCW} as
\begin{equation}
        T \epr = - \dt \mathcal{G} + \dot{\mathcal{W}}_{\mathrm{fuel}} \, ,
	\label{eq:epFW}
\end{equation}
where
\begin{equation}
	\begin{split}
		\mathcal{G} & = {\textstyle \sum_{X}} [X] \left( \mu_{X} - RT \right) + {\textstyle \sum_{Y}} [Y] \left( \mu_{Y} - RT \right) - \mu_{\W} L_{\W} \\
		 = &\cM \mu_{\M} + \cA \mu_{\A} + \cMs \left( \mu_{\Ms} - \mu_{\W} \right) + \cAs \left( \mu_{\As} - 2\mu_{\W} \right) + \cF \left( \mu_{\F} - \mu_{\W} \right) + \\
                   &- RT \left( \cM + \cA + \cMs + \cAs + \cF + \cW \right)
	\end{split}
	\label{}
\end{equation}
is a semigrand Gibbs potential, and
\begin{equation}
        \dot{\mathcal{W}}_{\mathrm{fuel}} := I_{\F} (\mu_{\F} - \mu_{\W}) \, .
	\label{DEF:FW}
\end{equation}
is the fueling chemical work per unit of time (\emph{i.e.}, the fueling power).
The derivation of Eq.~\eqref{DEF:FW} for an arbitrary CRN is discussed in Refs.~\cite{rao16:crnThermo,falasco18,rao18}.

If $\mu_{\F} = \mu_{\W}$, Eq.~\eqref{eq:epFW} shows that $\mathcal{G}$ is a monotonically decreasing function in time, given that $T \epr \geq 0$ by virtue of the second law of thermodynamics.
Its minimum value --- \emph{i.e.}, the equilibrium value --- under the constraint given by the conservation law (Eq.~\eqref{eq:EScl}) is found by minimizing the function $\Lambda = \mathcal{G} - \lambda L_{\M}$, where $\lambda$ is the Lagrange multiplier corresponding to $L_{\M}$.
The equilibrium concentrations thus satisfy the following conditions
\begin{equation}
	\begin{aligned}
		0 = \at{\der{\Lambda}{\cM}}{\mathrm{eq}} & = \mu^{\mathrm{eq}}_{\M} - \lambda = \mu^{\circ}_{\M} + RT \ln \cM_{\mathrm{eq}} - \lambda \, , \\
		0 = \at{\der{\Lambda}{\cA}}{\mathrm{eq}} & = \mu^{\mathrm{eq}}_{\A} - 2\lambda = \mu^{\circ}_{\A} + RT \ln \cA_{\mathrm{eq}} - 2\lambda \, , \\
		0 = \at{\der{\Lambda}{\cMs}}{\mathrm{eq}} & = \mu^{\mathrm{eq}}_{\Ms} - \mu_{\W} - \lambda = \mu^{\circ}_{\Ms} + RT \ln \cMs_{\mathrm{eq}} - \mu_{\W} - \lambda \, , \\
                0 = \at{\der{\Lambda}{\cAs}}{\mathrm{eq}} & = \mu^{\mathrm{eq}}_{\As} - 2\mu_{\W} - 2\lambda = \mu^{\circ}_{\As} + RT \ln \cAs_{\mathrm{eq}} - 2\mu_{\W} - 2\lambda \, .
	\end{aligned}
	\label{}
\end{equation}
The equilibrium semigrand Gibbs potential reads
\begin{equation}
	\begin{split}
		\mathcal{G}_{\mathrm{eq}} & = \lambda L_{\M} - RT \left( \cM_{\mathrm{eq}} + \cA_{\mathrm{eq}} + \cMs_{\mathrm{eq}} + \cAs_{\mathrm{eq}} + \cF_{\mathrm{eq}} + \cW_{\mathrm{eq}} \right) \\
		& = \cM \mu^{\mathrm{eq}}_{\M} + \cA \mu^{\mathrm{eq}}_{\A} + \cMs \left( \mu^{\mathrm{eq}}_{\Ms} - \mu_{\W} \right) + \cAs \left( \mu^{\mathrm{eq}}_{\As} - 2\mu_{\W} \right) + \\ 
                &   - RT \left( \cM_{\mathrm{eq}} + \cA_{\mathrm{eq}} + \cMs_{\mathrm{eq}} + \cAs_{\mathrm{eq}} + \cF_{\mathrm{eq}} + \cW_{\mathrm{eq}} \right)
				\, ,
	\end{split}
	\label{}
\end{equation}
which leads by direct calculation to equation~\ref{es_gibbs} in the main text:
\begin{equation}
	\mathcal{G} - \mathcal{G}_{\mathrm{eq}} = RT {\sum_{X}} \left[ [X] \ln \frac{[X]}{[X]_{\mathrm{eq}}} - [X] + [X]_{\mathrm{eq}} \right] \ge 0 \, .
	\label{}
\end{equation}
Therefore, when $\mu_{\F} = \mu_{\W}$, the quantity $\mathcal{G} - \mathcal{G}_{\mathrm{eq}}$ is a Lyapunov function for the open network relaxing to equilibrium.
When $\mathcal{F}_{\mathrm{fuel}} = \mu_{\F} - \mu_{\W} \neq 0$, the fueling chemical work in Eq.~\eqref{eq:epFW} does not vanish, and the system is prevented from reaching equilibrium.

Equation~\ref{es_work} in the main text is obtained by integrating Eq.~\eqref{eq:epFW} from time $t=0$ to a generic time $t$.
In our simulation of energy storage, we focused on the special case in which the system at time $t=0$ is at equilibrium ($\mathcal{F}_{\mathrm{fuel}} = 0$).

\subsection{Parameters}
\label{sec:ESsim}

With reference to the model in figure~\ref{FIG_scheme} of the main text, the following parameters were used for all the simulations:
\begin{table}[h!]
\caption{
    Parameters used for the energy storage model depicted in figure~\ref{FIG_scheme} of the main text.
    Values of the backward kinetic constants were obtain through Eq.~\eqref{eq:ldb} in order to assure thermodynamic consistency.
    Note that $\cW$ is kept fixed, while we used $\cF$ to tune $\mathcal{F}_\mathrm{fuel}$ in the various discussions.
}
\begin{tabular}{l|l}
 $k_\mathrm{+1F}$ & $5             \;  \text{M}^{-1} \text{s}^{-1}$ \\
 $k_\mathrm{+1W}$ & $1\cdot10^{-3} \;  \text{M}^{-1} \text{s}^{-1}$ \\
 $k_\mathrm{+2}$  & $1             \;  \text{M}^{-1} \text{s}^{-1}$ \\
 $k_\mathrm{+3F}$ & $1\cdot10^{-6} \;  \text{s}^{-1}$               \\
 $k_\mathrm{+3W}$ & $5 \;  \text{s}^{-1}$               \\
 $k_\mathrm{+4} $ & $1\cdot10^{-1} \;  \text{s}^{-1}$               
\end{tabular}
\qquad \qquad
\begin{tabular}{l|l}
 $ \mu^\circ_{\M}$    &  $-2\cdot10^{3}   \;  \text{J/mol}$ \\  
 $ \mu^\circ_{\Ms}$   &  $-3\cdot10^{3}   \;  \text{J/mol}$ \\  
 $ \mu^\circ_{\As}$ &  $-4\cdot10^{3}   \;  \text{J/mol}$ \\  
 $ \mu^\circ_{\A}$  &  $ 9\cdot10^{3}   \;  \text{J/mol}$ \\  
 $ \mu^\circ_\F$    &  $ 11\cdot10^{3}  \;  \text{J/mol}$ \\  
 $ \mu^\circ_\W$    &  $-11\cdot10^{3}  \;  \text{J/mol}$  
\end{tabular}
\qquad \qquad
\begin{tabular}{l|l}
 $ L_{\M}$    &  $1          \; \text{M}$ \\
 $[\text{W}]$ &  $1          \; \text{M}$ 
\end{tabular}
\label{tab:ESparameters}
\end{table}

\section{Dissipative Synthesis}
\label{sec:DS}

\subsection{Dynamics}
\label{sec:DSdynamics}

With the addition of the extraction mechanism, the evolution in time of the concentrations of the species $\M$, $\Ms$, $\A$, and $\As$ is ruled by the following rate equations
\begin{equation}
	\dt
	\underbrace{
		\begin{pmatrix}
			\cM \\
			\cMs \\
			\cAs \\
			\cA
		\end{pmatrix}
	}_{\textstyle [\bm{X}]}
	= 
	\underbrace{
		\begin{pmatrix}
			-1 	& -1	& 0 	& 0  	& 0 	& 2 \\
			1 	& 1		& -2 	& 0  	& 0 	& 0 \\
			0 	& 0		& 1 	& -1  	& -1 	& 0 \\
			0 	& 0		& 0 	& 1  	& 1 	& -1
		\end{pmatrix}
	}_{\textstyle \matSX}
        \cdot
	\underbrace{
		\begin{pmatrix}
			k_{+1\F} \cF \cM - k_{-1\F} \cMs \\
			k_{+1\W} \cW \cM - k_{-1\W} \cMs \\
			k_{+2} \cMs^{2} - k_{-2} \cAs \\
			k_{+3\F} \cAs - k_{-3\F} \cA \cF^2 \\
			k_{+3\W} \cAs - k_{-3\W} \cA \cW^2 \\
			k_{+4} \cA - k_{-4} \cM^{2} \\
		\end{pmatrix}
	}_{\textstyle \bm{J} = \bm{J}_{+} - \bm{J}_{-}}
	+
	\begin{pmatrix}
		2 I_{\mathrm{ext}} \\
		0 \\
		0 \\
		- I_{\mathrm{ext}}
	\end{pmatrix}
	\, ,
	\label{eq:DSrateX}
\end{equation}
where the current of extraction reads $I_{\mathrm{ext}} = k_{\mathrm{ext}} \cA$.

We examine this system at the steady state, in which all concentrations are stationary: $\dt [X]_{\mathrm{ss}} = 0$ for all $X$.
Their expressions are not analytical, but can be easily obtained numerically, thus showing that the steady state state is unique within a broad range of values that we examined.

\subsection{Thermodynamics}
\label{sec:DSthermo}

For the dissipative synthesis model at the steady state, Eq.~\eqref{eq:eprCW} becomes
\begin{equation}
	T \epr = \dot{\mathcal{W}}_{\mathrm{chem}} \, ,
	\label{eq:eprDSCW}
\end{equation}
where the chemical work per unit of time now reads
\begin{equation}
	\dot{\mathcal{W}}_{\mathrm{chem}} = \mu_{\F} I_{\F} + \mu_{\W} I_{\W} + 2 \mu_{\M} I_{\mathrm{ext}}  - \mu_{\A} I_{\mathrm{ext}} \, .
	\label{eq:DSCW}
\end{equation}

In order to construct the entropy balance as in equation~\ref{ds_power} of the main text, we once again need to consider conservation vectors~\eqref{eq:ESclL} and~\eqref{eq:ESclMMbGauge}, \emph{i.e.} a basis of the cokernel of $\matS$.
\begin{align}
	\bm \ell_{\M} & =
	\kbordermatrix{
		& \greyt{\M} & \greyt{\Ms} & \greyt{\As} & \greyt{\A} & \greyt{\F} & \greyt{\W} \\
		& 1 & 1 & 2 & 2 & 0 & 0
	} \, , \label{eq:DSclL} \\
	\bm \ell_{\W} & =
	\kbordermatrix{
		& \greyt{\M} & \greyt{\Ms} & \greyt{\As} & \greyt{\A} & \greyt{\F} & \greyt{\W} \\
		& 0 & 1 & 2 & 0 & 1 & 1
	} \, .  \label{eq:DSclMMbGauge}
\end{align}
Now, both of these vectors identify broken conserved quantities.
The former corresponds to the conserved quantity relative to the monomer
\begin{equation}
	L_{\M} = \bm \ell_{\M} \cdot
	\begin{pmatrix}
		[\bm{X}] \\
		[\bm{Y}]
	\end{pmatrix}
	= \cM + \cMs + 2 \cAs + 2 \cA \, .
	\label{}
\end{equation}
In the framework of Ref.~\cite{rao16:crnThermo}, this is a broken conservation law because of the presence of the extraction mechanism. 
Here its value does not change by construction of the model, since every $\A$ which is exchanged is readily replaced by 2 $\M$ molecules
\begin{equation}
	\dt L_{\M} = 2 I_{\mathrm{ext}} - 2 I_{\mathrm{ext}} = 0 \, .
	\label{}
\end{equation}
The latter represents the $\F$/$\W$ conservation law
\begin{equation}
	L_{\W} = \bm \ell_{\W} \cdot
	\begin{pmatrix}
		[\bm{X}] \\
		[\bm{Y}]
	\end{pmatrix}
	= \cMs + 2 \cAs + \cF + \cW \, ,
	\label{}
\end{equation}
which is broken by the fueling mechanism
\begin{equation}
	\dt L_{\W} = I_{\F} + I_{\W} \, .
	\label{eq:DSL}
\end{equation}

At the steady state all time derivative vanish, and we can use Eq.~\eqref{eq:DSL} to recast the chemical work per unit of time in Eq.~\eqref{eq:DSCW} into
\begin{equation}
	\dot{\mathcal{W}}_{\mathrm{chem}} = \dot{\mathcal{W}}_{\mathrm{fuel}} + \dot{\mathcal{W}}_{\mathrm{ext}}
	\label{eq:CW=FW+EW}
\end{equation}
where
\begin{equation}
	\dot{\mathcal{W}}_{\mathrm{fuel}} = I_{\F} \left( \mu_{\F} - \mu_{\W} \right) \, .
	\label{}
\end{equation}
is the input power, and
\begin{equation}
	\dot{\mathcal{W}}_{\mathrm{ext}} = I_{\mathrm{ext}} (2 \mu_{\M} - \mu_{\A})
	\label{}
\end{equation}
is the output power.
Combining Eq.~\eqref{eq:CW=FW+EW} with Eq.~\eqref{eq:eprDSCW}, we obtain equation~\ref{ds_power} of the main text.

\subsection{Supplementary plots for $-\dot{\mathcal{W}}_\mathrm{ext}$ and $\dot{\mathcal{W}}_\mathrm{fuel}$}
\label{sec:DSsim}

\begin{figure}[h!]
	\centering
	\includegraphics[width=.99\textwidth]{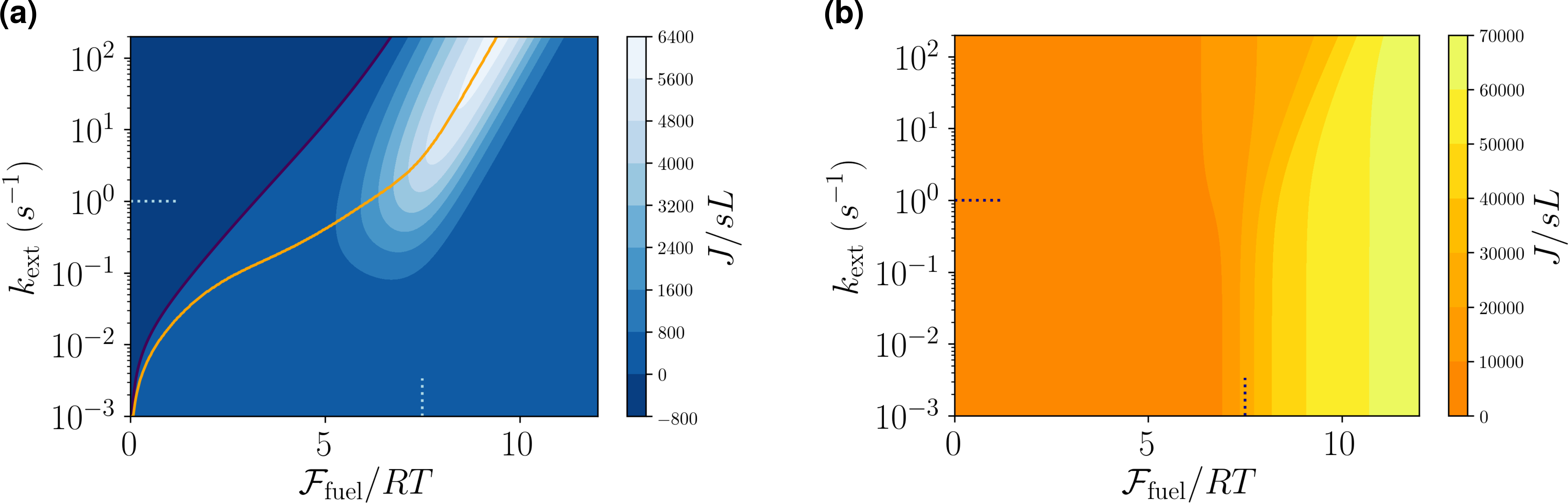}
	\label{fig:PW}
	\caption{
            \textbf{(a)} Minus the output power ($-\dot{\mathcal{W}}_\mathrm{ext}$) and \textbf{(b)} input power ($\dot{\mathcal{W}}_\mathrm{fuel}$) plotted in the same range of parameter as in figure~\ref{FIG_DS} of the main text.
                    The efficiency is given by the ratio of the two plots, according to equation~\ref{ds_eta} of the main text.
	}
\end{figure}


\subsection{Linear Regime}
\label{sec:linearRegime}

For $k_{\mathrm{ext}} = 0$ and $\mathcal{F}_\mathrm{fuel} = \mu_{\F} - \mu_{\W} = 0$, the entropy production at the steady state vanishes, and hence the steady state is an equilibrium one ($[X]_{\mathrm{eq}}$).
For
\begin{subequations}
	\begin{align}
		k_{\mathrm{ext}} & \ll 1 \\
                \mathcal{F}_\mathrm{fuel} = \mu_{\F} - \mu_{\W} & \ll RT \label{eq:LRconditionF}
	\end{align}
\end{subequations}
the entropy production is close to zero and hence the system is in a linear regime close to equilibrium.
In this regime, we can write the steady-state concentrations as $[X]_{\mathrm{ss}} = [X]_{\mathrm{eq}}(1+f_{X}/RT)$, where $f_{X} \ll RT$ for all $X$ encode the linear shifts from equilibrium.
Regarding the chemostatted ones, without loss of generality, we write $\cF = \cF_{\mathrm{eq}} (1 + \mathcal{F}_\mathrm{fuel}/RT)$ and $\cW = \cW_{\mathrm{eq}}$, where $\mu^\circ_{\F} + RT \ln \cF_{\mathrm{eq}} = \mu^\circ_{\W} + RT \ln \cW_{\mathrm{eq}}$.
In this way, when approximating the chemical potentials of the chemostats using the fact that $\mathcal{F}_\mathrm{fuel} \ll RT$, the equality in Eq.~\ref{eq:LRconditionF} is recovered.

By insterting the above expressions into the rate equations, Eq.~\eqref{eq:DSrateX} and \eqref{eq:ESrateY}, one obtains the analytical solution of the dissipative synthesis model at the steady-state in the linear regime.
Indeed, by discarding second order terms and exploiting the properties of the equilibrium state ($J_{+\rho}^\mathrm{eq} = J_{-\rho}^\mathrm{eq}$), the rate equations read
\begin{equation}
	\mathbb{M}_{\mathrm{X}}^{\mathrm{X}}
        \cdot
	\begin{pmatrix}
		f_{\M} \\
		f_{\Ms} \\
		f_{\As} \\
		f_{\A}
	\end{pmatrix}
	+  \mathcal{F}_\mathrm{fuel} \, \mathbb{M}_{\F}^{\mathrm{X}}
	= 
	\begin{pmatrix}
		2 I_{\mathrm{ext}} \\
		0 \\
		0 \\
		- I_{\mathrm{ext}}
	\end{pmatrix}
	\, 
	\quad \text{and} \quad \quad \quad
	\mathbb{M}_{\mathrm{X}}^{\F}
        \cdot
	\begin{pmatrix}
		f_{\M} \\
		f_{\Ms} \\
		f_{\As} \\
		f_{\A}
	\end{pmatrix}
	+ \mathcal{F}_\mathrm{fuel} \, \mathbb{M}_{\F}^{\F}
	= I_{\F}
	\, ,
	\label{eq:RLrateEq}
\end{equation}
for the internal and chemostatted species, respectively.
The extraction current is given by $I_{\mathrm{ext}} = k_{\mathrm{ext}} \cA_{\mathrm{eq}}$, while the matrix $\mathbb{M}$ is a 6 by 6 matrix which encodes both the topology and the kinetics of the linear regime dynamics
\begin{equation}
	\mathbb{M} := \matS \cdot \diag\left\{ \bm{J}^{\mathrm{eq}}_{+} \right\} \cdot \matS\transpose /RT \, ,
	\label{eq:matMXX}
\end{equation}
where
\begin{equation}
	\bm{J}^{\mathrm{eq}}_{+} = 
	\begin{pmatrix}
		k_{+1\F} \cF_{\mathrm{eq}} \cM_{\mathrm{eq}} &
		k_{+1\W} \cW_{\mathrm{eq}} \cM_{\mathrm{eq}} &
		k_{+2} \cMs_{\mathrm{eq}}^{2}  &
		k_{+3\F} \cAs_{\mathrm{eq}} &
		k_{+3\W} \cAs_{\mathrm{eq}} &
		k_{+4} \cA_{\mathrm{eq}} 
	\end{pmatrix}
	\label{eq:LRflux}
\end{equation}
are the equilibrium forward fluxes.
The labels $\mathrm{X}$ and $\F$ in Eq.~\eqref{eq:RLrateEq} select blocks of $\mathbb{M}$ corresponding to internal and fuel species, respectively, as sketched in the scheme below.
\begin{figure}[h!]
    \centering
\begin{tikzpicture}[
style1/.style={
  matrix of math nodes,
  every node/.append style={text width=#1,align=center,minimum height=5ex},
  nodes in empty cells,
  left delimiter=[,
  right delimiter=],
  },
]
\matrix[style1=0.65cm] (1mat)
{
  & & & & & \\
  & & & & & \\
  & & & & & \\
  & & & & & \\
  & & & & & \\
  & & & & & \\
};
\draw[loosely dashed]
  (1mat-4-1.south west) -- (1mat-4-6.south east);
\draw[loosely dashed]
  (1mat-5-1.south west) -- (1mat-5-4.south east);
\draw[loosely dashed]
  (1mat-1-4.north east) -- (1mat-6-4.south east);
\draw[loosely dashed]
  (1mat-1-5.north east) -- (1mat-4-5.south east);
\node[font=\huge] 
at ([xshift=-60pt]1mat-3-1.south west) {$\mathbb{M} =$};
\node[font=\LARGE] 
at ([yshift=-10pt]1mat-2-2.east) {$\mathbb{M}_\mathrm{X}^\mathrm{X}$};
\node[] 
at ([yshift=-10pt]1mat-2-5) {$\mathbb{M}_\mathrm{F}^\mathrm{X}$};
\node[] 
at ([yshift=-10pt]1mat-2-6) {$\mathbb{M}_\mathrm{W}^\mathrm{X}$};
\node[] 
at (1mat-5-2.east) {$\mathbb{M}_\mathrm{X}^\mathrm{F}$};
\node[] 
at (1mat-6-2.east) {$\mathbb{M}_\mathrm{X}^\mathrm{W}$};
\node[font=\large] 
at ([yshift=-10pt]1mat-5-5.east) {$\mathbb{M}_\mathrm{F,W}^\mathrm{F,W}$};
\node[] 
at ([xshift=-25pt]1mat-1-1.west) {$\greyt\M$};
\node[] 
at ([xshift=-25pt]1mat-2-1.west) {$\greyt\Ms$};
\node[] 
at ([xshift=-25pt]1mat-3-1.west) {$\greyt\A$};
\node[] 
at ([xshift=-25pt]1mat-4-1.west) {$\greyt\As$};
\draw[decoration={brace,mirror,raise=12pt},decorate]
  (1mat-1-1.north west) -- 
  node[left=15pt] {} 
  (1mat-4-1.south west);
\draw[decoration={brace,mirror,raise=12pt},decorate]
  (1mat-4-1.south west) -- 
  node[left=17pt] {$\greyt{\F}$} 
  (1mat-5-1.south west);
\draw[decoration={brace,mirror,raise=12pt},decorate]
  (1mat-5-1.south west) -- 
  node[left=17pt] {$\greyt{\W}$} 
  (1mat-6-1.south west);
\draw[decoration={brace,raise=7pt},decorate]
  (1mat-1-1.north west) -- 
  node[above=8pt] {$\greyt{\M} \enspace\quad \greyt{\Ms} \enspace\quad \greyt{\As} \enspace\quad \greyt{\A}$} 
  (1mat-1-4.north east);
\draw[decoration={brace,raise=7pt},decorate]
  (1mat-1-4.north east) -- 
  node[above=10pt] {$\greyt{\F}$} 
  (1mat-1-5.north east);
\draw[decoration={brace,raise=7pt},decorate]
  (1mat-1-5.north east) -- 
  node[above=10pt] {$\greyt{\W}$} 
  (1mat-1-6.north east);
\end{tikzpicture}
\, ,
\end{figure}

Let us now introduce the index ``$\mathrm{a}$'' to denote the activated species which are neither exchanged nor extracted ($\Ms$ and $\As$), whereas the index ``$\mathrm{e}$'' denotes the extracted/injected species ($\A$ and $\M$).
The rate equations can thus be further split into
\begin{equation}
	\begin{aligned}
		0 & = \mathcal{F}_\mathrm{fuel} \, \mathbb{M}^{\mathrm{a}}_{\F}  + \mathbb{M}^{\mathrm{a}}_{\mathrm{a}} \cdot \begin{pmatrix} f_{\Ms} \\ f_{\As} \end{pmatrix} + \mathbb{M}^{\mathrm{a}}_{\mathrm{e}} \cdot \begin{pmatrix} f_{\M} \\ f_{\A} \end{pmatrix} \\
		- I_{\mathrm{ext}} & = \mathcal{F}_\mathrm{fuel} \, \mathbb{M}^{\A}_{\F}  + \mathbb{M}^{\A}_{\mathrm{a}} \cdot \begin{pmatrix} f_{\Ms} \\ f_{\As} \end{pmatrix} + \mathbb{M}^{\A}_{\mathrm{e}} \cdot \begin{pmatrix} f_{\M} \\ f_{\A} \end{pmatrix} \\
		I_{\F} & = \mathcal{F}_\mathrm{fuel} \, \mathbb{M}^{\F}_{\F}  + \mathbb{M}^{\F}_{\mathrm{a}} \cdot \begin{pmatrix} f_{\Ms} \\ f_{\As} \end{pmatrix} + \mathbb{M}^{\F}_{\mathrm{e}} \cdot \begin{pmatrix} f_{\M} \\ f_{\A} \end{pmatrix} \, .
	\end{aligned}
	\label{eq:onsager0}
\end{equation}
We now observe that from the definition of conservation law, the following constraint holds
\begin{equation}
	\bm 0 = \mathbb{M} \, \bm{\ell}_{\M}\transpose = \mathbb{M}_{\mathrm{a}} \cdot \begin{pmatrix} 1 \\ 2 \end{pmatrix} + \mathbb{M}_{\mathrm{e}} \cdot \begin{pmatrix} 1 \\ 2 \end{pmatrix} \, ,
	\label{}
\end{equation}
which implies that
\begin{equation}
	\mathbb{M}_{\M} = - 2 \mathbb{M}_{\A} - \mathbb{M}_{\mathrm{a}} \cdot \begin{pmatrix} 1 \\ 2 \end{pmatrix} \, .
	\label{}
\end{equation}
This allows us to rewrite Eqs.~\eqref{eq:onsager0} as
\begin{equation}
	\begin{aligned}
		0 & = \mathcal{F}_\mathrm{fuel} \, \mathbb{M}^{\mathrm{a}}_{\F}  + \mathbb{M}^{\mathrm{a}}_{\mathrm{a}} \cdot \begin{pmatrix} f_{\Ms} - f_{\M} \\ f_{\As} - 2 f_{\M} \end{pmatrix} + \left( f_{\A} - 2 f_{\M} \right) \, \mathbb{M}^{\mathrm{a}}_{\A}  \\
		- I_{\mathrm{ext}} & = \mathbb{M}^{\A}_{\F} \, \mathcal{F}_\mathrm{fuel} + \mathbb{M}^{\A}_{\mathrm{a}} \cdot \begin{pmatrix} f_{\Ms} - f_{\M} \\ f_{\As} - 2 f_{\M} \end{pmatrix} + \left( f_{\A} - 2 f_{\M} \right) \, \mathbb{M}^{\A}_{\A}  \\
		I_{\F} & = \mathcal{F}_\mathrm{fuel} \, \mathbb{M}^{\F}_{\F}  + \mathbb{M}^{\F}_{\mathrm{a}} \cdot \begin{pmatrix} f_{\Ms} - f_{\M} \\ f_{\As} - 2 f_{\M} \end{pmatrix} + \left( f_{\A} - 2 f_{\M} \right) \, \mathbb{M}^{\F}_{\A} \, .
	\end{aligned}
	\label{eq:onsager1}
\end{equation}
We now solve the first of the three equations above for the vector in parenthesis, using the fact that $\mathbb{M}^{\mathrm{a}}_{\mathrm{a}}$ is nonsingular.
\begin{equation}
	\begin{pmatrix} f_{\Ms} - f_{\M} \\ f_{\As} - 2 f_{\M} \end{pmatrix} = - (\mathbb{M}^{\mathrm{a}}_{\mathrm{a}})^{-1} \cdot \left[ \mathcal{F}_\mathrm{fuel} \, \mathbb{M}^{\mathrm{a}}_{\F}  + \left( f_{\A} - 2 f_{\M} \right) \, \mathbb{M}^{\mathrm{a}}_{\A}  \right] \, .
	\label{}
\end{equation}
This follows from the fact that $\mathbb{M}^{\mathrm{a}}_{\mathrm{a}}$ is Gramian \cite{horn85}, and $\matS^{\mathrm{a}}$ contains linearly independent vectors.
Therefore, the last two equations in \eqref{eq:onsager1} can be recast into
\begin{equation}
	\begin{aligned}
		- I_{\mathrm{ext}} & = \mathcal{F}_\mathrm{fuel} \, \left[ \mathbb{M}^{\A}_{\F} - \mathbb{M}^{\A}_{\mathrm{a}} \cdot (\mathbb{M}^{\mathrm{a}}_{\mathrm{a}})^{-1} \cdot \mathbb{M}^{\mathrm{a}}_{\F} \right] + \left( f_{\A} - 2 f_{\M} \right) \, \left[ \mathbb{M}^{\A}_{\A} - \mathbb{M}^{\A}_{\mathrm{a}} \cdot (\mathbb{M}^{\mathrm{a}}_{\mathrm{a}})^{-1} \cdot \mathbb{M}^{\mathrm{a}}_{\A} \right]  \\
		I_{\F} & = \mathcal{F}_\mathrm{fuel} \, \left[ \mathbb{M}^{\F}_{\F} - \mathbb{M}^{\F}_{\mathrm{a}} \cdot (\mathbb{M}^{\mathrm{a}}_{\mathrm{a}})^{-1} \cdot \mathbb{M}^{\mathrm{a}}_{\F} \right] + \left( f_{\A} - 2 f_{\M} \right) \, \left[ \mathbb{M}^{\F}_{\A} - \mathbb{M}^{\F}_{\mathrm{a}} \cdot (\mathbb{M}^{\mathrm{a}}_{\mathrm{a}})^{-1} \cdot \mathbb{M}^{\mathrm{a}}_{\A} \right] \, .
	\end{aligned}
	\label{eq:onsager2}
\end{equation}
Changing signs conveniently, we can rewrite the above equations in terms of the Onsager matrix $\mathbb{L}$, which expresses the linear dependence of currents from forces when the system is close to equilibrium:
\begin{align}
    \left(
    \begin{array}{c}
      I_F \\
      I_\mathrm{ext}
    \end{array}
    \right)
    = \mathbb{L}
    \left(
    \begin{array}{c}
      \mu_\mathrm{F} - \mu_\mathrm{W} \\
      2\mu_\mathrm{M} - \mu_\mathrm{A_2}
    \end{array}
    \right) \, .
    \label{lin_reg}
\end{align}
Indeed, in the linear regime the chemical force associated to the extraction currents is $2 \mu_{\M} - \mu_{\A} = 2 f_{\M} - f_{\A}$.
The entries of the Onsager matrix are given by
\begin{equation}
	\mathbb{L} =
	\begin{pmatrix}
		\mathbb{M}^{\F}_{\F} - \mathbb{M}^{\F}_{\mathrm{a}} \cdot (\mathbb{M}^{\mathrm{a}}_{\mathrm{a}})^{-1} \cdot \mathbb{M}^{\mathrm{a}}_{\F} & \mathbb{M}^{\F}_{\mathrm{a}} \cdot (\mathbb{M}^{\mathrm{a}}_{\mathrm{a}})^{-1} \cdot \mathbb{M}^{\mathrm{a}}_{\A} - \mathbb{M}^{\F}_{\A} \\
		\mathbb{M}^{\A}_{\A} - \mathbb{M}^{\A}_{\mathrm{a}} \cdot (\mathbb{M}^{\mathrm{a}}_{\mathrm{a}})^{-1} \cdot \mathbb{M}^{\mathrm{a}}_{\A} & \mathbb{M}^{\A}_{\mathrm{a}} \cdot (\mathbb{M}^{\mathrm{a}}_{\mathrm{a}})^{-1} \cdot \mathbb{M}^{\mathrm{a}}_{\F} - \mathbb{M}^{\A}_{\F} 
	\end{pmatrix}
        :=
	\begin{pmatrix}
            \mathbb{L}_{11} & \mathbb{L}_{12} \\
	    \mathbb{L}_{21}& \mathbb{L}_{22} 
	\end{pmatrix}.
	\label{LR:Lmatrix}
\end{equation}

We can use Eq.~\eqref{lin_reg} to analytically evaluate the efficiency $\eta_\mathrm{ds}$ introduced in equation~\ref{ds_eta} of the main text, as well as the output power $\dot{\mathcal{W}}_\mathrm{ext}$, in terms of $k_\mathrm{ext}$ and $\mathcal{F}_\mathrm{fuel}$, namely the control parameters in the model:
\begin{align}
    \eta_\mathrm{ds} = -\frac{I_\mathrm{ext}(I_\mathrm{ext} - \mathcal{F}_\mathrm{fuel}\mathbb{L}_{12})}{\mathcal{F}_\mathrm{fuel}(I_\mathrm{ext}\mathbb{L}_{12} + \mathcal{F}_\mathrm{fuel} \mathrm{det}[\mathbb{L}])} \, ; \quad
    \dot{\mathcal{W}}_\mathrm{ext} = \frac{I_\mathrm{ext}(I_\mathrm{ext} - \mathcal{F}_\mathrm{fuel}\mathbb{L}_{12})}{\mathbb{L}_{11}} \, .
\end{align}
When $\mathcal{F}_\mathrm{fuel}$ is kept fixed, the values of $k_\mathrm{ext}$ which extremise $\eta_\mathrm{ds}$ and $-\dot{\mathcal{W}}_\mathrm{ext}$ are readily found by deriving the previous expressions and look for the unique stable points:
\begin{align}
    \mathrm{max\ efficiency: }& \quad k^*_\mathrm{ext} = \frac{\sqrt{\mathbb{L}_{11}\mathbb{L}_{22}\mathrm{det}[\mathbb{L}]} - \mathrm{det}[\mathbb{L}]}{\mathbb{L}_{12}\cA_\mathrm{eq}}\mathcal{F}_\mathrm{fuel} \label{maxeff}  \\
    \mathrm{max\ output\ power: }& \quad k^*_\mathrm{ext} = \frac{\mathbb{L}_{12}}{2\cA_\mathrm{eq}}\mathcal{F}_\mathrm{fuel} \, . \label{maxpow} 
\end{align}
The above equations define the sets of points of maximum efficiency and efficiency at maximum power for any value of $\mathcal{F}_\mathrm{fuel}$ within the linear regime.
By equating the right hand sides of Eq.~\eqref{maxeff} and Eq.~\eqref{maxpow}, one obtains that these two expressions coicide if and only if $\mathbb{L}_{12} = \mathbb{L}_{21} = 0$, which is never the case for coupled currents.

When evaluated using the parameters in SI\_Table~\ref{tab:ESparameters}, Eq.~\eqref{LR:Lmatrix} reads
\begin{equation}
	\mathbb{L} =
	\begin{pmatrix}
	    17.7835 & 3.74893 \\
            3.74893 & 23.7732
	\end{pmatrix}
	\cdot 10^{-8} \, mol^2/sLJ
	\label{}
\end{equation}
where the cross coefficients are equal according to the Onsager reciprocal relations.

When the analytical solution is plotted against $k_{\mathrm{ext}}$ and $\mathcal{F}_\mathrm{fuel}$, we obtain the plot in SI\_Fig.~\ref{fig:LR}b, where both maximum efficiency and efficiency at maximum power are highlighted as in figure~\ref{FIG_DS} of the main text.
An enlargement of the linear region of figure~\ref{FIG_DS} of the main text is shown in SI\_Fig.~\ref{fig:LR}a.

\begin{figure}[h!]
	\centering
	\includegraphics[width=\textwidth]{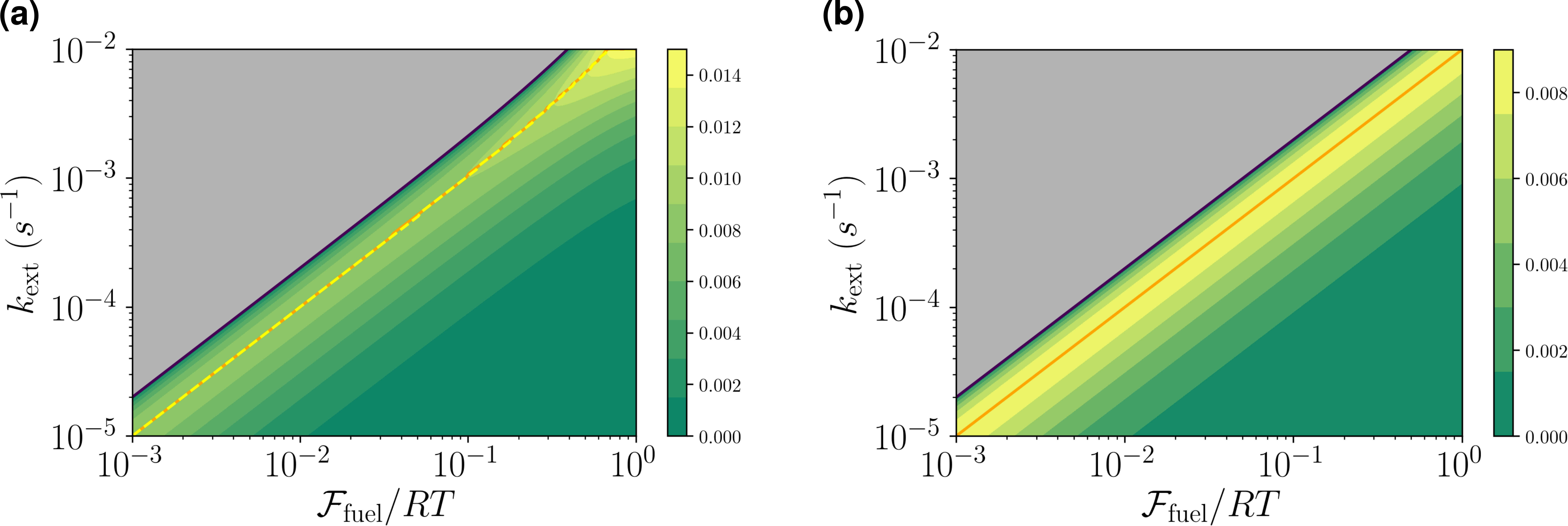}
	\caption{
                Comparison between exact simulation of the full dynamics \textbf{(a)} and analytical formula obtained in the linear regime \textbf{(b)} for the efficiency in the linear regime.
                The log scale emphasizes the changes of magnitude of these values.
                For low forces and extraction rates --- where Eq.~\eqref{lin_reg} is a good approximation --- the two plots clearly coincide.
                When $\mathcal{F}_\mathrm{fuel}$ is of the order of $0.1$ (in units of $RT$) and $k_{\mathrm{ext}}$ reaches $10^{-3} \, s^{-1}$ differences in both numerical values and shape emerge.
                In particular, we see that the increase in efficiency visible for high $\mathcal{F}_\mathrm{fuel}$ and $k_{\mathrm{ext}}$ in (a) is a genuine nonequilibrium feature as it is absent in the linear regime, (b).
    }
	\label{fig:LR}
\end{figure}

\vspace{10pt}
\end{widetext}


%

\end{document}